\documentclass[pra,a4paper,groupedaddress,showkeys,showpacs]{revtex4}
\usepackage{graphicx}
\usepackage{tabularx}
\usepackage{amsmath}
\usepackage{amsfonts}
\usepackage{amssymb}
\usepackage{natbib}

\begin{document}
\title{Analysis of surface waves generated on subwavelength-structured silver films}
\date{\today}
\author{G. Gay}\author{O. Alloschery}\author{J.
Weiner}\email{jweiner@irsamc.ups-tlse.fr} \affiliation{IRSAMC/LCAR\\
Universit\'e Paul Sabatier, 118 route de Narbonne,\\31062
Toulouse, France}
\author{H. J. Lezec} \affiliation{Thomas J. Watson Laboratories of
Applied Physics, California Institute of Technology, Pasadena,
California 91125 USA}\affiliation{Centre National de la Recherche
Scientifique, 3, rue Michel-Ange, 75794 Paris cedex 16, France}
\author{C. O'Dwyer}\affiliation{Tyndall National Institute, University College Cork, Cork, Ireland
}\author{M. Sukharev}\author{T. Seideman}\affiliation{Department
of Chemistry, Northwestern University, 2145 Sheridan Road,
Evanston, Illinois 60208-33113 USA}

\keywords{plasmon; surface wave; nanostructure}

\begin{abstract}
Using transmission electron microscopy (TEM) to analyse the
physical-chemical surface properties of subwavlength structured
silver films and finite-difference time-domain (FDTD) numerical
simulations of the optical response of these structures to
plane-wave excitation, we report on the origin and nature of the
persistent surface waves generated by a single slit-groove motif
and recently measured \cite{GAV06a} by far-field optical
interferometry. The surface analysis shows that the silver films
are free of detectable oxide or sulfide contaminants, and the
numerical simulations show very good agreement with the results
previously reported.
\end{abstract}

\pacs{42.25.Fx. 73.20.Mf. 78.67.-n}

\maketitle

\section{Introduction}
The optical response of subwavelength-structured metallic films
has enjoyed a resurgence of interest in the past few years due to
the quest for an all-optical solution to the inexorable drive for
ever-smaller, denser integrated devices operating at ever-higher
band width \cite{IPRA-Nano}.  Two basic questions motivate
research in this field: how to confine micron-sized light waves to
subwavelength dimensions \cite{DSA06} and how to transmit this
light without unacceptable loss over at least tens of microns
\cite{ZSB05,BDE03}.  Although surface waves and in particular
periodic arrays of surface plasmon polaritons (SPPs) have received
a great deal of attention as promising vehicles for subwavelength
light confinement and transport, detailed understanding of their
generation and early time evolution (within the first few optical
cycles) in and on real metal films is still not complete
\cite{XZM04,ZMM06,QL02,LHR05}. Recent measurements
\cite{GAV06a,GAV06b} of far-field interference fringes arising
from surface wave generation in single slit-groove structures on
silver films have characterized the amplitude, wavelength and
phase of the surface waves.  After a rapid amplitude decrease
within the first three microns from the generating groove, waves
persisting with near-constant amplitude over tens of microns were
observed. Such long-range transport is the signature of a "guided
mode" SPP, but the measured wavelength was found to be markedly
shorter than the expectation from conventional theory
\cite{Raether}.  One possible reason advanced for the disparity
between experiment and theory was the presence of an oxide or
sulfide dielectric layer on the silver surface, and it has been
suggested \cite{LH06} that an 11~nm layer of silver sulfide would
bring experiment and theory into agreement.  We report here the
results of two investigations: one experimental, into the
physical-chemical surface properties of the silver structures with
the intent of detecting the presence of a dielectric layer and the
other theoretical, into the calculated optical response of the
untarnished silver slit-groove structures using the FDTD technique
to numerically solve Maxwell's equations. These studies shown no
evidence of oxide or sulfide layers on the silver surface and the
numerical solutions to Maxwell's equations show good agreement
with measurements reported in \cite{GAV06a}.

\section{TEM analysis of structured silver surfaces}  Three
typical structured samples were chosen for examination with
fabrication dates of about 12 months, 6 months, and 1 week from
the date of the TEM analysis.  The structures dating from 6 months
and 12 months were part of the series of structures actually used
in the previous reports \cite{GAV06a,GAV06b}.  The subwavelength
structures consist of a 400~nm silver film evaporated onto a fused
silica substrate (Corning 7980 UV grade$^{\copyright}$ 25~mm square, 1~mm
thick optically polished on both sides to a roughness of no more
than 0.7~nm). The structured substrates are stored in
Fluoroware$^{\circledR}$ sample holders, 25~mm diameter and 1~mm
in depth at the center.

\subsection{Methods} Electron transparent sections for cross-sectional transmission electron
microscopy (TEM) examination were prepared by sample thinning to
electron transparency using standard focused Ga$^{+}$-ion beam
(FIB) milling procedures \cite{milling} in a FEI 200 FIB
workstation and placed on a holey carbon support. The TEM
characterization was performed using a Philips CM300 Schottky
field emission gun (FEGTEM) microscope operating at 300~kV. The
field emission electron source is ideal for applications requiring
high coherency, high brightness at high magnification, or small
focused probes of selected areas. The FEGTEM has a point
resolution of 0.2~nm and an information limit of 0.12~nm. The
minimum focused electron beam probe size is 0.3~nm. In dark field
images, one or more diffracted beams are allowed to pass the
objective aperture while the direct straight-through beam is
blocked by the aperture. In contrast to the direct beam, the
diffracted beam interacts strongly with the specimen and selection
of a particular diffracted beam allows better visual phase
differentiation.

\subsection{Examination of silver (Ag) layers} 
\begin{figure}
\centering
\includegraphics[width=0.5\columnwidth]{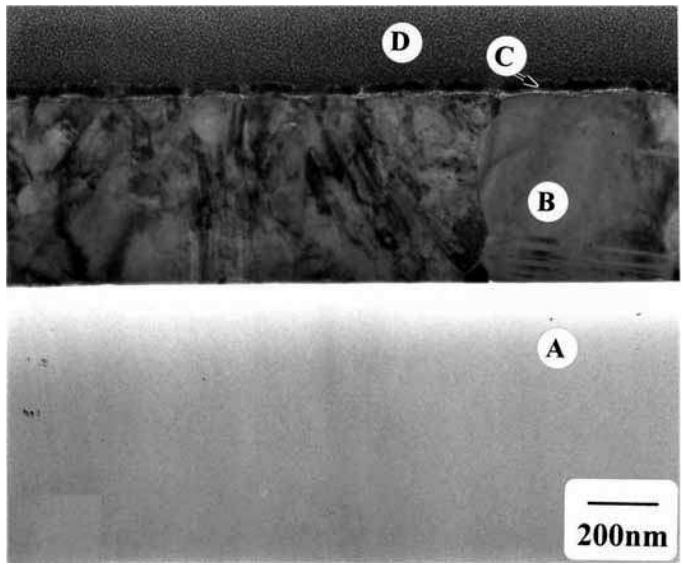}
\caption{Cross-sectional bright-field through-focal TEM micrograph
of the Ag layer on a fused silica substrate. Labels refer to:
fused silica substrate \textbf{A}, silver layer \textbf{B},
capping gold layer \textbf{C}, and capping platinum layer
\textbf{D}.  The darkened features in the Ag layer (\textbf{B})
are dislocations and grain boundaries.}\label{ColmLRTEM}
\end{figure}

Typical regions of Ag layer that were
thinned to electron transparency are shown at relatively low
magnification in the bright field micrograph in
Fig.\,\ref{ColmLRTEM}. An immediate distinction can be made
between the fused silica substrate (\textbf{A}) and the Ag deposit
(\textbf{B}). Protective capping layers of Au and Pt, applied at
the time of analysis, are marked at (\textbf{C}) and (\textbf{D})
respectively. The latter are used to prevent any `top down' ion
damage of the cross-section during the ion beam thinning
preparation.

\begin{figure}
\begin{minipage}{0.4\columnwidth}
\centering
\includegraphics[width=\columnwidth]{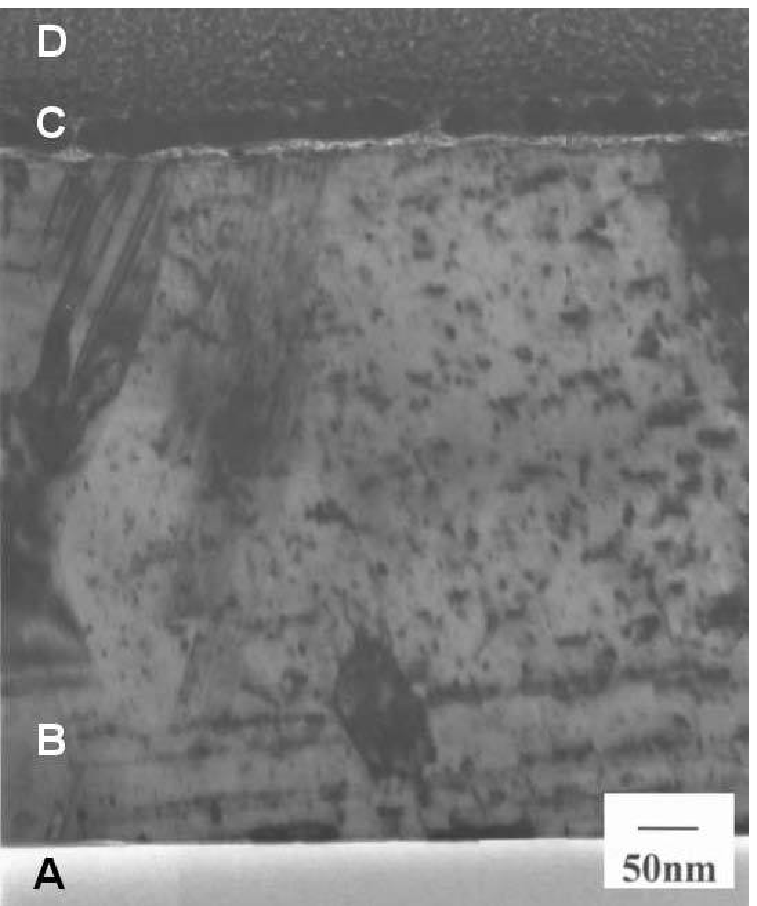}\caption{Cross-sectional bright-field through-focal TEM micrograph of the Ag
layer (\textbf{B}) on the fused silica substrate (\textbf{A}). The
protective capping layers of Au and Pt are marked at (\textbf{C})
and (\textbf{D}) respectively. }\label{ColmHRTEMa}
\end{minipage}\hspace{0.02\columnwidth}
\begin{minipage}{0.4\columnwidth}
\centering
\includegraphics[width=\columnwidth]{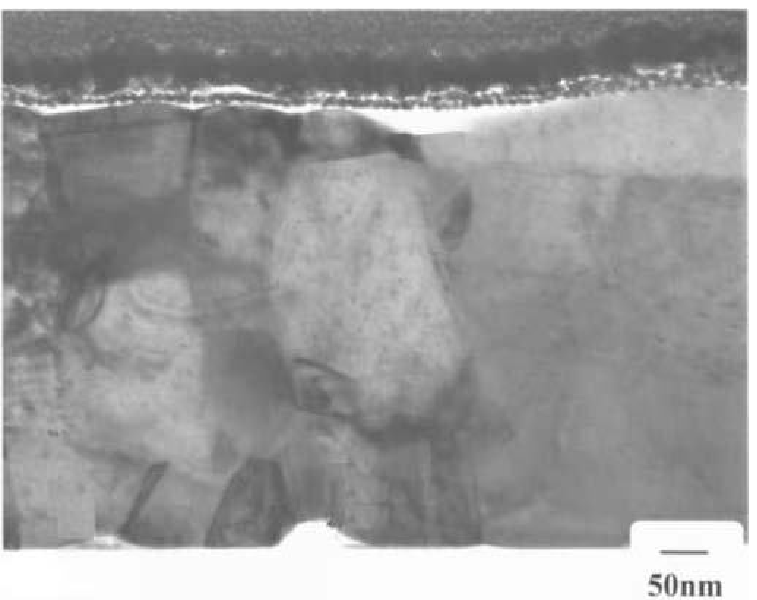}\caption{A higher
magnification of the upper surface showing the Au
particle-containing protection layer. The undulations and voids
indicate a surface nonuniformity somewhat greater than that of the
samples in general ($\pm\sim 10$~nm from 400~nm nominal thickness)
but the site was chosen to more vividly show the initial line of
gold capping nanoparticles.}\label{ColmHRTEMb}
\end{minipage}
\end{figure}

The white colored line marked by an arrow in Fig.\,\ref{ColmLRTEM}
and shown in more detail in Fig.\,\ref{ColmHRTEMa} is a band that
contains very fine particles of Au that are formed during the
initial stages of the deposition of this metal.

The fine particles do not exhibit the same degree of absorption
contrast as the bulk of the Au formed above them because the
particle diameters are less than the thickness of the
cross-sectional slice and thus do not extend through the full
thickness of the TEM sample, as do the relatively coarse grains
constituting the bulk of the Au protection layer. A higher
magnification TEM micrograph of this region is shown for clarity
in Fig.\,\ref{ColmHRTEMb}. It is clearly observed that the Au
nanoparticles form a separate layer above the Ag deposit and any
local undulations in the Ag layer are observed to be devoid of any
oxide or sulfide layer.

\begin{figure}
\centering
\includegraphics[width=1.0\columnwidth]{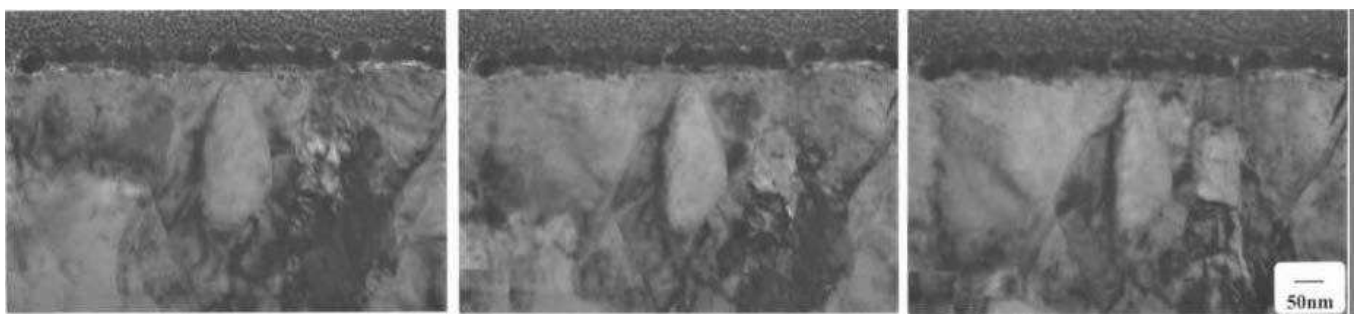}
\caption{Cross-sectional bright-field through-focal TEM
micrographs of the Ag layer on the fused silica substrate. The
images were acquired at over-focus (left), in-focus (middle) and
under-focus (right).}\label{ColmFresnel}
\end{figure}

Fresnel contrast methods were used to examine the upper surface
regions of the silver in more detail and a typical area is shown
in Fig.\,\ref{ColmFresnel}.
Imaging of layers in cross-section, in which the electrons travel
parallel to the interfaces between the two materials, always
results in Fresnel fringes. Because the amplitude and phase
changes that occur when an electron is scattered elastically are
characteristic of the atomic number, there will be changes in the
elastic scattering directly related to the form of the projected
scattering potential when there is a composition change at an
interface viewed in projection \cite{wang}, analogous to a phase
grating filtering of the scattered electronic wave function.
Interface Fresnel effects can thus provide a signature on the form
and magnitude of any compositional discontinuity that is present.
The visibility of these fringes depends on the thickness of the
specimen and on the defocus value of the microscope. The contrast
of the fringes is higher with increasing defocus, at both positive
and negative values from the minimum contrast condition.  From
Fig.\,\ref{ColmFresnel}, the critical observation is in the
\emph{absence} of Fresnel effects at the surface of the metal as a
function of focus conditions other than at the localized regions
associated with the gold protective coating. The lack of Fresnel
fringes at the upper surface indicates a lack of variation in the
scattering potentials and hence of the chemistry or the
composition at the interface. The lack of Fresnel signature
demonstrates that silver surface is not covered by a detectable
layer of sulfide or oxide since their presence would exhibit
differences in scattering potential by comparison with the metal
itself\,\cite{fultz}.  The lower detection limit for such
compositional differences is a few tenths nanometer layer
thickness.

\subsection{Discussion of the surface analysis}  The high
resolution TEM analysis failed to detect the presence of oxide or
sulfide coating on any of the structured silver samples.  This
result is in fact not surprising and is consistent with a simple
estimate of the surface chemical impurity likely to be present.
Since the structures are kept in air-tight sample holders, the
total,static volume of air to which they are exposed during
storage is $\simeq 2000$~mm$^3$.  The trace fractional
concentrations of sulfur dioxide and hydrogen sulfide in ordinary
laboratory air is less than $0.2\times 10^{-9}$\,\cite{BPB69} and
therefore about $10^{10}$ impurity molecules are available to
react with the 625~mm$^{2}$ silver surface.  The average linear
distance between impurity molecules, assuming all of them reacted
with the surface, is therefore about 250~nm -- a surface density
below the detection limit of the FEGTEM instrument.

\begin{figure}
\centering
\includegraphics*[width=0.5\columnwidth]{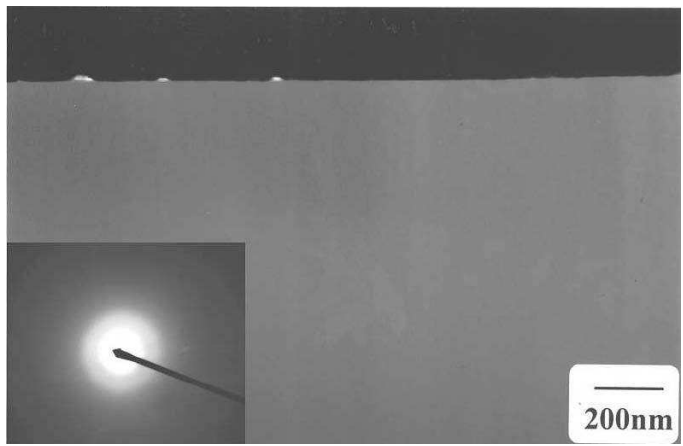}
\caption{Cross-sectional bright-field through-focal TEM
micrographs of the tarnished Ag layer on the fused silica
substrate after more than two years exposed to ambient laboratory
conditions. Insert: electron diffraction pattern of the tarnished
Ag layer showing amorphous (noncrystalline) characteristics by
means of broad, diffuse diffraction rings. }\label{ColmTarnish}
\end{figure}

For comparison, TEM was also conducted on a similar Ag structure
left fully exposed to ambient laboratory environment for more than
two years. The resulting micrograph and electron diffraction
pattern are shown in Fig.\,\ref{ColmTarnish}. The tarnished Ag
layer is observed to be uniform and the corresponding electron
diffraction data shows a fully amorphous (noncrystalline)
character evidenced by the broad diffuse rings. The detection of a
relatively high intensity signal from a nominally nondiffracting
material yields valuable information on not only the crystallinity
of the material, but also the chemical composition. Comparing the
electron diffraction data for the tarnished samples to that for
the silver structures described earlier, we find no evidence of a
similar noncrystalline material on the surface of the untarnished
Ag shown in Fig.\,\ref{ColmLRTEM}.

\section{Calculation of the far-field transmission}

\begin{figure}
\includegraphics[width=0.8\columnwidth]{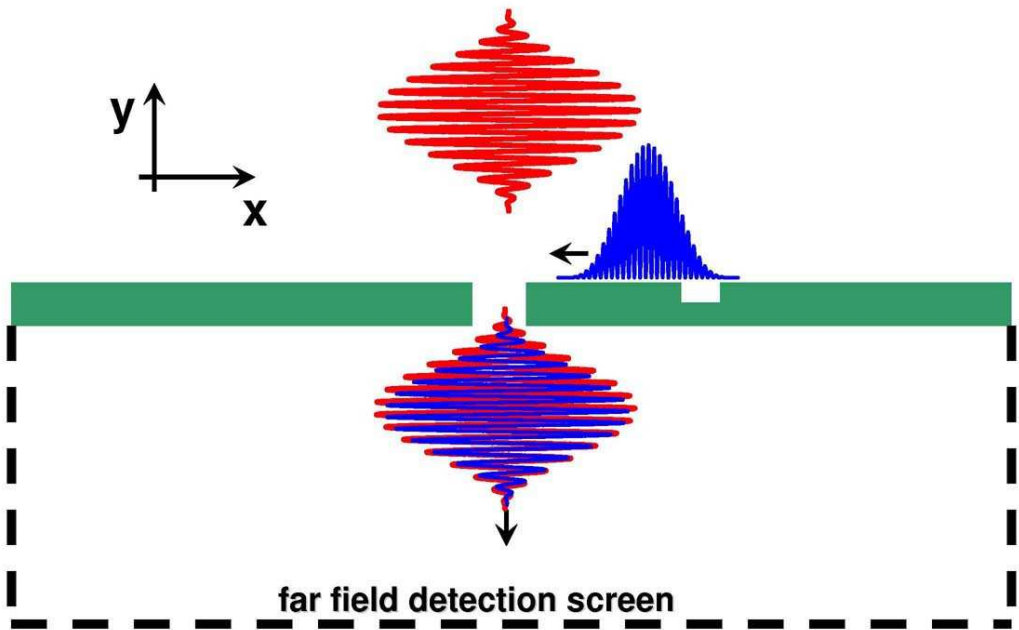}
\caption{Schematic structure and detection contour for FDTD
simulations. The slit and the groove widths are 100~nm, the groove
depth is 100~nm and the silver film thickness is 400~nm.
Interference of waves directly propagating through the slit with
waves originating from the groove and propagating to the slit as
surface waves give rise to the simulated interference pattern
shown in Fig.\,\ref{results}}\label{contour}
\end{figure}

The optical response of structured metal surfaces is simulated
using a finite-difference-time-domain (FDTD)
approach\,\cite{Taflove2000}. The subwavelength slit-groove
structures\,\cite{GAV06a,GAV06b} can be modelled in two
dimensions, and we consider here the case of transverse magnetic
TM polarization (with the $H$-field transverse to the plane of
incidence and parallel to the long axis of the slit-groove
structures). The set-up for the simulations is shown in
Fig.\,\ref{contour}.
The Maxwell equations are thus,
\begin{eqnarray}
\varepsilon\frac{\partial E_{x}}{\partial t}&=&\frac{\partial
H_{z}}{\partial y}-J_{x},\nonumber\\
\varepsilon\frac{\partial E_{y}}{\partial t}&=&-\frac{\partial
H_{z}}{\partial x}-J_{y},\label{Maxwell equations}\\
\mu_{0}\frac{\partial H_{z}}{\partial t} &=&\frac{\partial
E_{x}}{\partial y}-\frac{\partial E_{y}}{\partial x}\nonumber
\end{eqnarray}
where $E_{x}$, $E_{y}$, and $H_{z}$ are the Cartesian components
of the electric and magnetic fields, $J_{x}$ and $J_{y}$ are the
Cartesian components of the current density $\vec{J}$,
$\varepsilon$ is the electric permittivity and $\mu_{0}$ is the
magnetic permeability of free space. In metallic regions of space
$\varepsilon$ is a complex valued, frequency-dependent function.
Within the standard Drude model \cite{Huffman1983} it is given as,
\begin{equation}
\varepsilon\left(\omega\right)=\varepsilon_{0}\left(\varepsilon_{\infty}-\frac{\omega_{p}^{2}}{\omega^{2}+i\Gamma\omega}
\right),\label{Drude profile}
\end{equation}
where $\varepsilon_{0}$ is the electric permittivity of free
space,
$\varepsilon_{\infty}=\varepsilon\left(\omega\rightarrow\infty\right)$
is the dimensionless infinite frequency limit of the dielectric
constant, $\omega_{p}$ is the bulk plasmon frequency, and $\Gamma$
is the damping rate. In order to properly describe the material
dispersion of the metal structures we numerically fit the real and
imaginary parts of the Drude dielectric constant in the form of
Eq.\,\ref{Drude profile} to the experimental data collected in
\cite{LynchHunter}. The Drude parameters obtained in the
wavelength regime ranging from $750$ nm to $900$ nm for silver are
$\varepsilon_{\infty}=3.2938$, $\omega_{p}=1.3552\times10^{16}$
rad s$^{-1}$, and $\Gamma=1.9944\times10^{14}$ rad s$^{-1}$, which
correspond to $\operatorname{Re}\left [\varepsilon\right]
=-33.9767$ and $\operatorname{Im}\left[\varepsilon\right] =3.3621$
at $\lambda_0=852$~nm. These fitted parameters are close to those
determined in \cite{GAV06a} by ellipsometry,
$\operatorname{Re}\left[\varepsilon\right]=-33.27$ and
$\operatorname{Im}\left[\varepsilon\right]=1.31$.  Numerical
results were not sensitive to this range of parameter variability.

The frequency dependence of the dielectric constant
Eq.\,\ref{Drude profile} results in an additional equation to the
Maxwell equations\,(\ref{Maxwell equations}), describing the time
evolution of the current density $\vec{J}$ in the metallic region
of space\,\cite{Ziolkowski1995}.
\begin{equation}
\frac{\partial\vec{J}}{\partial
t}=\alpha\vec{J}+\beta\vec{E},\label{current}
\end{equation}
where $\alpha=-\Gamma$, $\beta=\varepsilon_{0}\omega_{p}^{2}$ and
$\varepsilon=\varepsilon_{0}\varepsilon_{\infty}$. In the
surrounding free space $\varepsilon=\varepsilon_{0}$, and
$\alpha=\beta=0$ and hence $\vec{J}$ vanishes \cite{Gray2003PRB}.

As a reference for the optical response of the Drude metal we also
implement perfect electric conductor (PEC) boundary conditions,
where the metal dielectric constant is set to negative infinity,
and all electromagnetic field components are therefore strictly
zero in metal regions.

The light source is a plane wave, incident perpendicular to the
metal surface and depending on time as,
\begin{equation}
E_{\text{inc}}\left(t\right)=E_{0}f\left(t\right)\cos\omega t,
\label{incident source}
\end{equation}
where $E_{0}$ is the peak amplitude of the pulse, $f\left(t\right)
=\sin^{2}\left(\pi t/\tau\right)$ is the pulse envelope, and
$\tau$ is the pulse duration ($f\left( t>\tau\right)=0$.

Time propagation of Eqs.\,\ref{Maxwell equations},\,\ref{current}
is performed by a leapfrogging technique \cite{Taflove2000}. In
order to prevent nonphysical reflection of outgoing waves from the
grid boundaries, we employ perfectly matched layer (PML) absorbing
boundaries, within which $H_{z}$ is split into two additive
subcomponents \cite{BerengerPML}.  To avoid accumulation of
spurious electric charges at the ends of the excitation line and
generate a pure plane wave with a well-defined incident
wavelength, we embed the ends in the PML regions. We have tested
this approach by direct comparison with the total field/scattered
field technique\,\cite{Taflove2000}. In all simulations the
spatial grid ranges from $x=-16.2$\,$\mu$m to $16.2$\,$\mu$m and
$y=-1.79$\,$\mu$m to $1.79$\,$\mu$m. Convergence is achieved with
a spatial step size of $\delta x=\delta y\lesssim4$\,nm and a
temporal step size of $\delta t=\delta x/\left(2c\right)$, where
$c$ denotes the speed of light in vacuum.

All simulations have been performed on distributed memory parallel
computers at the National Energy Research Scientific Computing
Center and San Diego Supercomputer Center. The parallel technique
used in our simulations is described in detail in
\cite{SukharevJCP2006}.

\subsection{Numerical results}

\begin{figure}
\includegraphics*[width=0.65\columnwidth]{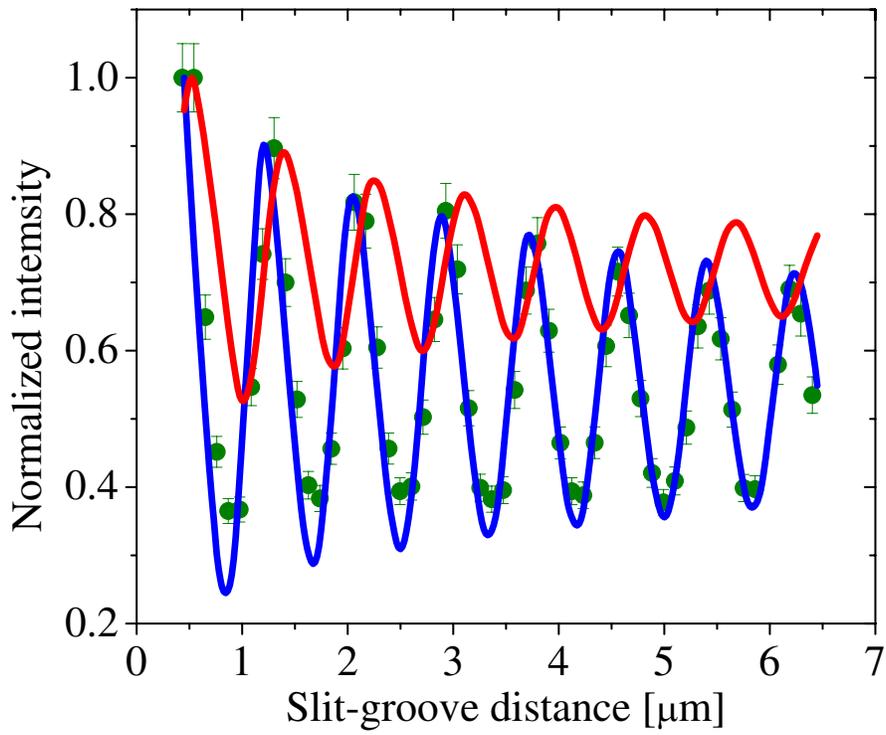}\caption{Comparison of FDTD simulation and experiment. Green points
are experimental data taken from\,\cite{GAV06a}.  Blue curve is
the FDTD result for the Drude model.  Red curve is the FDTD result
for the PEC model.}\label{results}
\end{figure}

Calculations of the intensity,
$I\sim\left(E_{x}^{2}+E_{y}^{2}+H_{z}^{2}\right)$, are performed
along a box contour as shown in Fig. \ref{contour}, including a
line parallel to the metal surface and located at the distance
$L_{D}$. The collected data is averaged over time and the spatial
coordinates for a range of slit-groove distances. Finally, the
space- and time-averaged intensity is normalized to unit maximum.
Our results converge with $L_{D}\gtrsim 1$\,$\mu$m and are
invariant to the pulse duration (hence converged to the CW limit)
with incident pulse durations $\tau\gtrsim 200$~fs.  A direct
comparison of the experimental data with both the PEC and the
Drude models is shown in Fig.\,\ref{results}.

\begin{figure}
\begin{minipage}[t]{0.48\columnwidth}
\centering
\includegraphics[width=1.0\columnwidth]{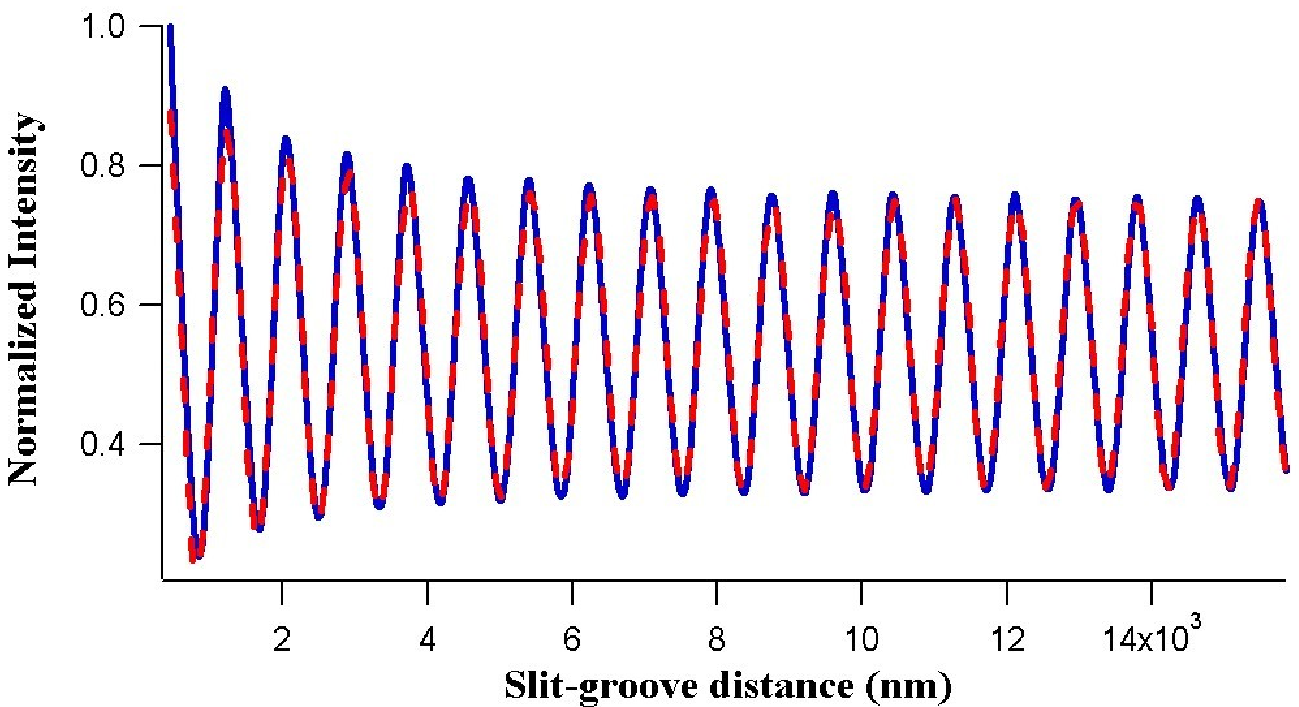}\caption{Blue curve plots the same Drude model FDTD calculation as in Fig.\,\ref{results}
but extended to 16~$\mu$m slit-groove distance.  Red dashed curve
plots the the analytic fitting function of Eq.\,\ref{damping
sin}.}\label{extended-Drude-PEC}
\end{minipage}\hspace{0.02\columnwidth}
\begin{minipage}[t]{0.45\columnwidth} \centering
\includegraphics[width=1.0\columnwidth]{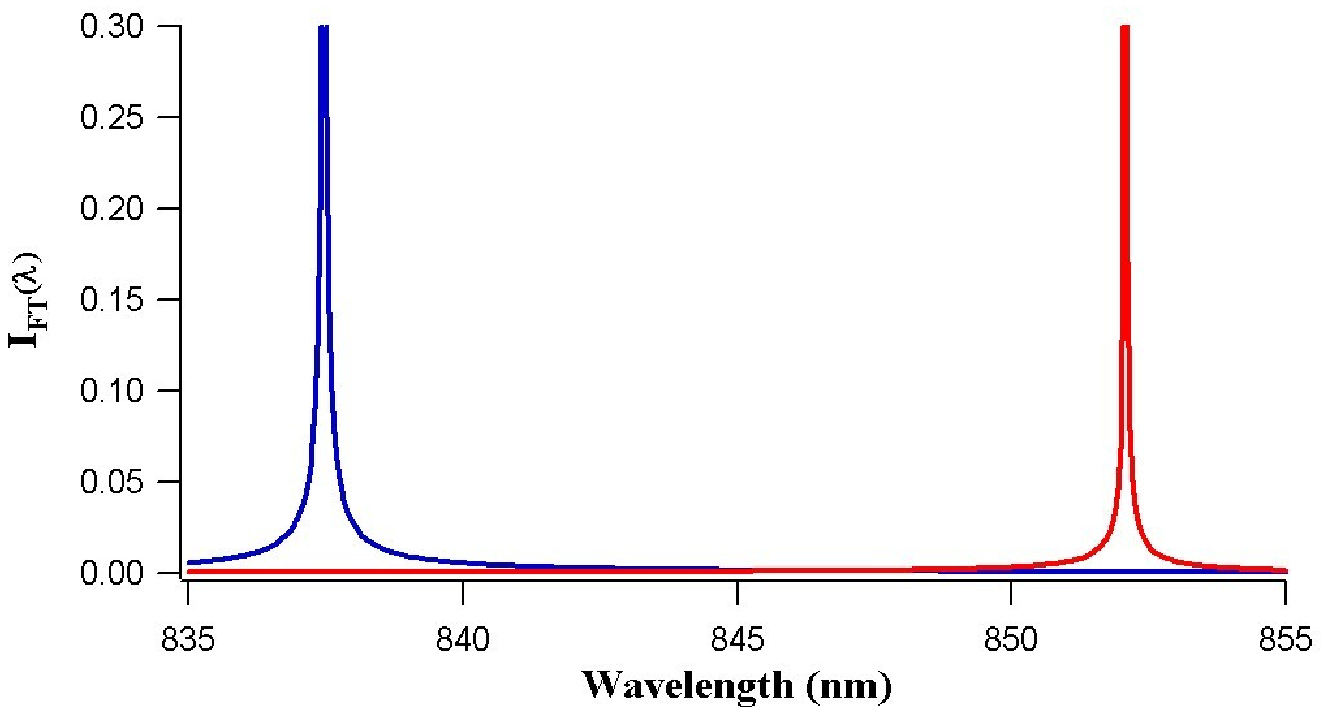}\caption{Fourier transform spectra of Eq.\,\ref{damping sin}
for curves fitted to the Drude (blue) and PEC (red) results of
FDTD simulations. Fitting parameters used in Eq.\,\ref{damping
sin} for Drude (PEC) models are, $A_1$:0.20686\,(0.061126),
$A_2$:0.19597\,(0.22834), $A_3$:-0.0005213\,(-0.00046489),
$A_4$:0.007374\,(0.0075026), $A_5$:3.3098\,(2.0663),
$A_6$:0.54336\,(0.71776).}\label{FT-spectrum}
\end{minipage}
\end{figure}

Clearly, the Drude model agrees very well with the data, whereas
the PEC model predicts oscillations of the intensity with a
noticeably larger wavelength and smaller amplitude. In order to
properly determine the wavelength of oscillations we use a $\cos$
function with an exponentially decreasing amplitude component and
constant offset to fit the data shown in Fig.\,\ref{results} and
extend both the FDTD simulation and the fitting function to a
slit-groove distance of 16~$\mu$m, well beyond the range of
experimental data. We then take the Fourier transform of the
fitted function over this extended range to obtain the associated
power spectrum expressed as a wavelength distribution. The results
are shown in Fig.\,\ref{FT-spectrum}.
\begin{equation}
I_{\mathrm{fit}}(x)=[A_1+A_2\exp(A_3x)]\cos(A_4x+A_5)+A_6
\label{damping sin}
\end{equation}
The FDTD simulation and analytic fit for larger slit-groove
distances are shown in Fig.\,\ref{extended-Drude-PEC} for the case
of the Drude model. The fitting function Eq.\,\ref{damping sin}
tracks reasonably well the FDTD results over the entire
slit-groove distance range. Figure\,\ref{FT-spectrum} shows the
normalized power spectrum corresponding to Eq.\,\ref{damping sin}
as a function of the wavelength. The two power spectrum plots of
Fig.\,\ref{FT-spectrum} provide the effective wavelength,
$\lambda_{\mathrm{eff}}$, at which surface waves propagate as well
as the distribution of modes around the peak. Note that the Drude
model for silver shows a marked blue shift in peak wavelength and
a noticeable broadening of the distribution compared to the
perfect metal PEC model.  The Drude model results in
$\lambda_{\mathrm{eff}}=837.482$~nm , whereas PEC model gives
$\lambda_{\mathrm{eff}}=852.066$~nm, very close to the free-space
reference wavelength $\lambda_0=852$~nm. The effective surface
index of refraction
\begin{equation}
n_{\mathrm{eff}}=\frac{\lambda_{\mathrm{inc}}}{\lambda
_{\mathrm{eff}}},\label{effective index}
\end{equation}
where $\lambda_{\mathrm{inc}}$ is the incident wavelength (in the
experiment $\lambda_{\mathrm{inc}}=852$ nm), leads to the
following values: $n_{\mathrm{eff}}\left(\text{Drude}\right)
=1.0173$ and $n_{\mathrm{eff}}\left(\text{PEC}\right) =0.9999$.

\section{Summary and conclusions}

Surface analysis by transmission electron microscopy of
subwavelength structured silver films used to investigate their
optical response\,\cite{GAV06a, GAV06b} showed no detectable
evidence of material on the surface other than silver. The
suggestion that an 11~nm sulfide layer may be present so as to
bring the interference pattern calculated by the authors of
Ref.\,\cite{LH06} into agreement with experiment is therefore not
confirmed.

In contrast to the calculations reported by Ref.\,\cite{LH06},
numerical solution of Maxwell's equations reported here using an
FDTD approach does show very good agreement with the fringe
amplitude and wavelength over the slit-groove range of the data
reported in \cite{GAV06a}. Extrapolation of the FDTD simulations
beyond the range of the measurements shows that the initially
decreasing amplitude of the fringe settles to an oscillation with
near-constant amplitude and fringe contrast.  These features
resemble the calculations of \cite{LH06}, but the results reported
here show much better agreement with experiment without the need
to invoke an 11~nm silver sulfide layer.  Fourier analysis of the
FDTD simulations reveal that the most probable wavelength in the
Fourier distribution is 837~nm, within a about 2 nanometers of the
expected long-range SPP wavelength of 839~nm. The reasons for the
discrepancies between the results of Ref.\,\cite{LH06}, which
employ a different approach using rigorous-coupled-wave-analysis
(RCWA) and those reported here have not yet been identified.

\begin{acknowledgments}
Support from the Minist{\`e}re d{\'e}l{\'e}gu{\'e} {\`a} l'Enseignement sup{\'e}rieur et {\`a}
la Recherche under the programme
ACI-``Nanosciences-Nanotechnologies," the R{\'e}gion Midi-Pyr{\'e}n{\'e}es
[SFC/CR 02/22], and FASTNet [HPRN-CT-2002-00304]\,EU Research
Training Network, is gratefully acknowledged as is support from
the Caltech Kavli Nanoscience Institute and from the AFOSR under
Plasmon MURI FA9550-04-1-0434. This research used resources of the
National Energy Research Scientific Computing Center, which is
supported by the Office of Science of the U.S. Department of
Energy under Contract No. DE-AC03-76SF00098, and San Diego
Supercomputer Center under Grant No. PHY050001. Discussions with
P. Lalanne, M. Mansuripur, and H. Atwater as well as computational
assistance from J. Yelk and Y. Xie are also gratefully
acknowledged.
\end{acknowledgments}


\begin{thebibliography}{99}
\bibitem{GAV06a} G. Gay, O. Alloschery, B. Viaris de Lesegno, C. O'Dwyer, J.
Weiner and H. J. Lezec, Nature Physics \textbf{2}, 262-267 (2006).

\bibitem{IPRA-Nano} See for example the Technical Digest of the Integrated
PHotonics Research and Application (IPRA) and Nanophotonics (Nano)
topical meetings of the Optical Society of America, Uncasville,
Connecticut USA, April 24-28, 2006 (ISBN 1-55752-807-1).

\bibitem{DSA06} J. A. Dionne, L. A. Sweatlock, H. A. Atwater and A. Polman,
Phys. Rev. B \textbf{73}, 035407-1--9 (2006) and references cited
therein.

\bibitem{ZSB05} R. Zia, M. D. Selker, and M. L. Brongersma, Phys. Rev. B
\textbf{71} 16543-1--9 (2005).

\bibitem{BDE03} W. L. Barnes, A. Dereux, and T. W. Ebbesen, Nature (London)
\textbf{424}, 824-830 (2003) and references cited therein.

\bibitem{XZM04} Y. Xie, A. Z. Zakharian, J. V. Moloney, and M. Mansuripur,
Opt. Express \textbf{13} 4485-4491 (2005).

\bibitem{ZMM06} A. R. Zakharian, J. V. Moloney, and M. Mansuripur, \textit{%
Surface plasmon polaritons on metallic surfaces}, unpublished.

\bibitem{QL02} Q. Cao and P. Lalanne, Phys. Rev. Lett. \textbf{88}, 057403
(2002).

\bibitem{LHR05} P. Lalanne, P. Hugonin, and J. C. Rodier, Phys. Rev. Lett.
\textbf{95}, 263902 (2005).

\bibitem{GAV06b} G. Gay, O. Alloschery, B. Viaris de Lesegno, J. Weiner, and
H. Lezec, Phys. Rev. Lett. \textbf{96}, 213901-1--4 (2006).

\bibitem{Raether} H. Raether,\textit{Surface Plasmons on Smooth and Rough
Surfaces and on Gratings}, (Springer-Verlag, Berlin, 1988).

\bibitem{LH06} P. Lalanne and J. P. Hugonin, Nature
Physics \textbf{2}, 556 (2006).

\bibitem{milling} L. A. Giannuzzi and F. A. Stevie, \emph{Micron.} \textbf{30%
}, 197 (1999).

\bibitem{fultz} B. Fultz and J. Howe, \textit{Transmission Electron
Microscopy and Diffractometry of Materials}, 2$^{nd}$ ed.,
Springer NY (2005).

\bibitem{wang} Z. L. Wang, \textit{Elastic and Inelastic Scattering in
Electron Diffraction and Imaging}, Plenum Press, NY (1995).

\bibitem{BPB69} H. E. Bennett, R L. Peck, D.K. Burge, and J. M.
Bennett, J. Appl. Phys. \textbf{40}, 3351-3360 (1969).

\bibitem{Taflove2000} A. Taflove and S. C. Hagness, \textit{Computational
Electrodynamics: The Finite-Difference Time-Domain Method} (Artech
House, Boston, 2000).

\bibitem{Huffman1983} C. F. Bohren and D. R. Huffman, \textit{Absorption and
Scattering of Light by Small Particles} (Wiley, New York, 1983).

\bibitem{LynchHunter} D. W. Lynch and W. R. Hunter, \textit{Handbook of
Optical Constants of Solids} (Academic, Orlando, 1985).

\bibitem{Ziolkowski1995} J. B. Judkins and R. W. Ziolkowski J. Opt. Soc. Am.
A \textbf{12}, 1974 (1995).

\bibitem{Gray2003PRB} S. K. Gray and T. Kupka, Phys. Rev. B \textbf{68},
045415 (2003).

\bibitem{BerengerPML} J.-P. Berenger, J. Comput. Phys. \textbf{114}, 185
(1994).

\bibitem{SukharevJCP2006} M. Sukharev and T. Seideman, J. Phys. Chem.
\textbf{124}, 144707 (2006).
\end{thebibliography}
\end{document}